\documentclass[prl,twocolumn]{revtex4}
\usepackage{amsmath}
\usepackage{amssymb}
\usepackage[dvips]{graphicx}
\usepackage{dcolumn}
\usepackage{color}
\graphicspath{{images/}}
\begin{document}

\title{Kibble-Zurek mechanism in topologically non-trivial zigzag chains of polariton micropillars}
\author{D. D. Solnyshkov}
\author{A. V. Nalitov}
\author{G. Malpuech}

\affiliation{Institut Pascal, PHOTON-N2, Universit\'{e} Clermont Auvergne, CNRS, 4 Avenue Blaise Pascal, 63178 Aubi\`{e}re Cedex, France.}

\begin{abstract}
We consider a zigzag chain of coupled micropillar cavities, taking into account the polarization of polariton states.
We show that the TE-TM splitting of photonic cavity modes yields topologically protected polariton edge states.
During the strongly non-adiabatic process of polariton condensation, the Kibble-Zurek mechanism leads to a random choice of polarization, equivalent to dimerization of polymer chains.
We show that dark-bright solitons appear as domain walls between polarization domains, analogous to the Su-Schrieffer-Heeger solitons in polymers. The soliton density scales as a power law with respect to the quenching parameter.
\end{abstract}

%\pacs{}
\maketitle

As initially shown by Kibble \cite{Kibble1976} for the expansion and cooling of the early Universe, and then for liquid helium by Zurek \cite{Zurek1985}, a system undergoing a second-order phase transition on a finite timescale develops domains with independent order parameters.
The Kibble-Zurek mechanism (KZM) allows predicting the typical size of the domains and, therefore, the densities of the topological defects on their boundaries.
Their scaling as a function of the quench rate is given by a power law, with the critical exponent of the transition being determined by its universality class \cite{Zurek1996}. 

A very relevant system to study the KZM are the quantum fluids such as atomic Bose-Einstein condensates (BECs) formed by cooling.
Indeed, quantum fluids support topological defects \cite{LeggettBook,VolovikReview}, their most famous example being a quantum vortex, which, contrary to a classical vortex, like a tornado, cannot disappear "by itself", via a continuous transformation.
This property is due to the difference in vortex and ground state topologies, which is guaranteed by the irrotational nature of the fluid described by a complex wavefunction \cite{Pitaevskii}.
The non-adiabatic cooling of such a fluid allows the development of topological defects obeying the KZM scaling, as confirmed by experiments \cite{Lamporesi2013} and also predicted for multi-component BECs \cite{Damski2010,Sabbatini2011}.
However, solitons in 1D and half-vortices in 2D spinor BECs \cite{LeggettBook,VolovikReview,Rubo2007} are only quasi-topological defects.
Indeed, a dark soliton transforms into a grey and eventually disappears by simple acceleration, and a half-vortex can be unwound by a divergent magnetic field.
Another system of interest to study the KZM are cavity exciton-polariton quantum fluids \cite{Microcavities,Carusotto2013rev}.
Because of the finite polariton lifetime, polariton condensation can be an out-of-equilibrium process driven by the condensation kinetics rather than by thermodynamics \cite{Kasprzak2008,Carusotto2013rev}.
As previously pointed out\cite{Matuszewski2014,Liew2015}, the establishment of a steady state by non-resonant pumping in an initially empty system cannot be an adiabatic process and is therefore equivalent to a quenching of the parameters of the system, leading to the appearance of topological defects.

Another class of systems which possess topologically protected states are periodic lattices with topologically non-trivial band structures characterized by non-zero Chern numbers, or Zak phase, depending on their dimensionality.
The most well-known examples of such systems are the topological insulators \cite{Hasan2010}, Kitaev chains supporting topologically protected Majorana states \cite{Kitaev2001} and dimer chains \cite{Su1980}.
Indeed, depending on the difference of the tunneling coefficients within and between the dimers, such chains form topologically different conduction bands, characterized by a $\pi$ difference in the Zak phase \cite{Zak1989}.
As it was shown recently \cite{Delplace2011}, the number of states in the conduction band depends on this phase, and the states which are not included in the bulk are localized on the edges.
These edge states do not rely on inter-particle interactions but are topologically protected: they are robust against disorder and perturbations.
Different implementations of topologically non-trivial band structures have been studied theoretically and experimentally in various systems, including photonics \cite{Schomerus2013,Lu2014}, optomechanics \cite{Schmidt2015,Peano2015}, excitons \cite{Zhou2014}, and plasmonic zigzag chains \cite{Poddubny,Poddubny2015,Slobozhanyuk2015,Sinev2015}.
Optical systems offer an important advantage compared to the electronic ones and to atomic BECs, because of the facility of their fabrication and the complete accessibility of the wavefunction in time, real and reciprocal space. Polaritonic systems were shaped as molecules and lattices \cite{Yamamoto2013,Cerda2010,Jacqmin2014}. Schemes for creating polariton topological insulators have been proposed \cite{Karzig2015,Nalitov2014b, Liew2014,Zhou2015}. While KZM, as a universal mechanism, has already been widely studied in various systems sharing some common properties with our proposal, such as zigzag ionic chains \cite{delCampo2010,Ulm2013,Pyka2013,delCampo2014}, where the phase transition and the topology correspond to physical arrangement of atoms, none of these possess the same key ingredients. 
A topologically non-trivial polaritonic chain therefore appears as an ideal system to study the complex interplay of topological ordering and KZM \cite{Bermudez2009}.

In this work, we describe polariton BEC in a zigzag chain of polariton micropillars with photonic spin-orbit coupling (SOC) \cite{Nalitov2014,Nalitov2014b,Sala2015}.
As a result, the polariton band is characterized by a non-zero Zak phase and the chain supports topologically protected edge states. These results are also valid for Rashba SOC present in atomic condensates \cite{Lin2}.
We show that with a focused non-resonant excitation spot condensation occurs on the edge states with polarization determined by the Zak phase.
When the system is excited homogeneously, a gas of dark-bright solitons (a spinor version of Su-Schrieffer-Heeger (SSH) solitons) is formed via the KZM. Here, the $k^2$ TE-TM SOC is crucial, allowing a homogeneous condensate, contrary to $k$-linear Rashba SOC.
We demonstrate that the soliton density follows a power law with respect to the quenching parameter - pumping intensity. The scaling exponent for different system parameters is close to 1/4, in agreement with the mean-field KZM theory for a 1D system.

\emph{Existence of topological edge states.} We first consider a zigzag chain of coupled 0D modes neglecting the spin, as usually done in theoretical analysis of the electronic dimer chains.
Let us call the first pillar in the chain "a", and let it be "below" the second pillar "b", so that the first link is oriented at 45$^\circ$ (see Fig. 1), which we will call "diagonal" direction (D), while the perpendicular direction shall be "anti-diagonal" (A, 135$^\circ$).
The pair "ab" forms the unit cell.
Following the definitions established in the previous works \cite{Delplace2011}, the tunneling constant in the first link (within the cell) is called $t'$, while the tunneling in the second link (between cells) is $t$.
The corresponding tight-binding Hamiltonian reads ($m$ is the cell number):
\begin{equation}
\hat{H}=\sum_{m}t'\hat{b}_{m}^{\dagger}\hat{a}_m+t\hat{a}_{m+1}^{\dagger}\hat{b}_{m}+H.c.
\end{equation}
where $\hat{a}$ and $\hat{b}$ operators act on the corresponding pillars.

Let us now consider that these 0D modes are constituted by photonic micro-pillars obtained by etching a planar cavity \cite{Ferrier2011}.
Each pillar ground state has two polarizations, which we assume to be degenerate.
On the other hand, the optical eigenmodes of the cavity are TE and TM polarized and have different effective masses \cite{Panzarini1999}. This makes the tunneling coefficients polarization-dependent \cite{Nalitov2014,Nalitov2014b,Sala2015} and different for the polarizations oriented longitudinal and transverse with respect to the link.
We therefore have $t<t'$ for D-polarization, for which the first link (labeled by $t'$) is longitudinal (Fig. 1(a,d)), and $t>t'$ for the A-polarization, for which the same link is transverse (Fig. 1(b,e)).
The relative difference in the longitudinal and transverse tunneling coefficients for typical parameters of a polariton micropillar lattice can be of the order of 10\% \cite{Nalitov2014b}.
This difference of the tunneling coefficients is equivalent to the dimerization of polymer chains, but associated with the polarization of the states.
The corresponding dimers are shown with black dashed lines in Fig. 1.

\begin{figure}[t]\label{fig1}
\includegraphics[scale=0.49]{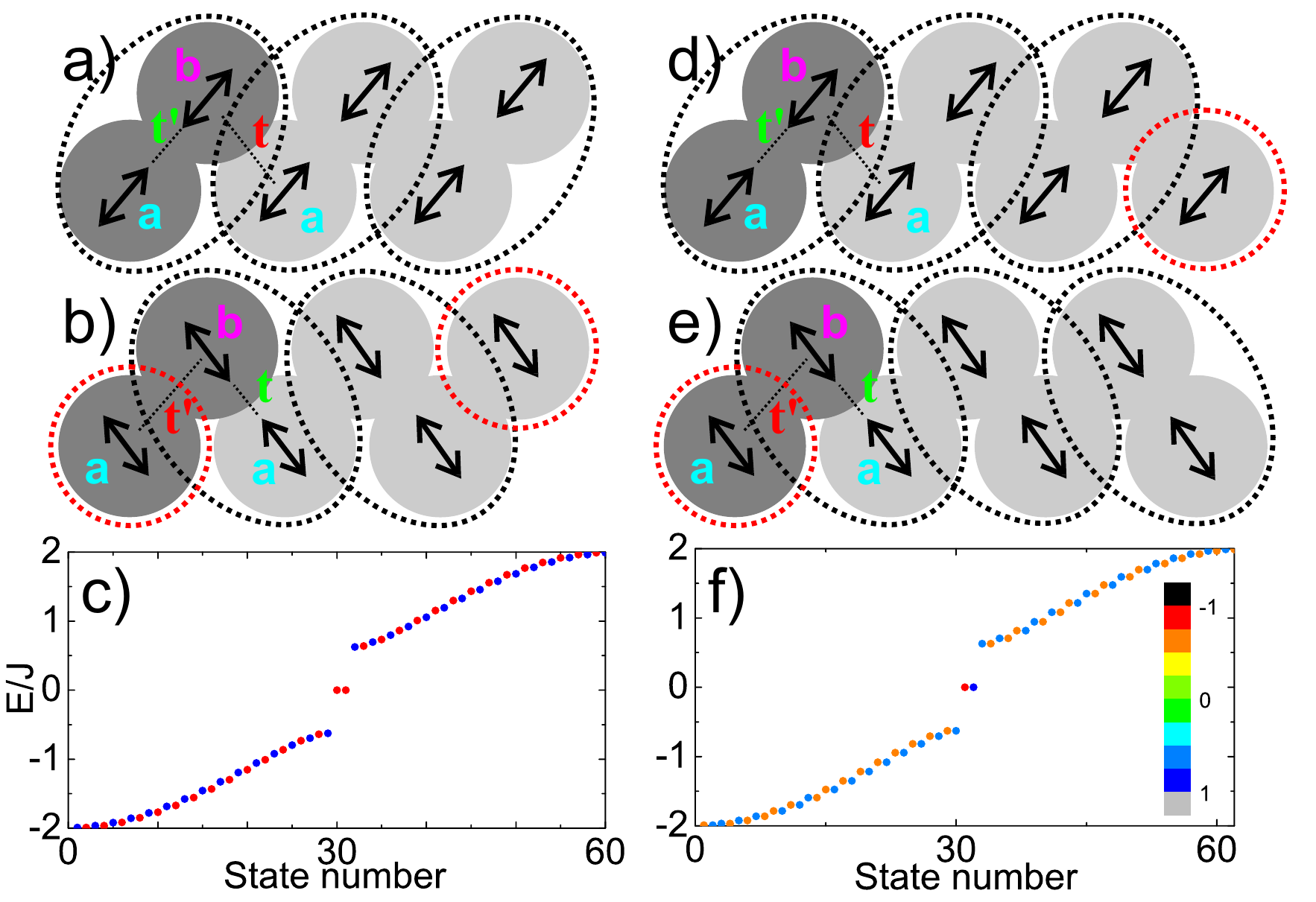} 
\caption{(color online) A scheme of a zigzag chain with even and odd number of  polariton pillars: "a" and "b" form a unit cell marked in dark grey. a,d) Diagonal polarization: $t'>t$, b,e) Anti-diagonal polarization: $t'<t$. c,f) Energy band of 30(c) and 31(f) pillar chains obtained from the tight binding Hamiltonian. The color shows the diagonal polarization degree of the states.}
\end{figure} 

For an even number of pillars, one can directly apply the SSH theory developed for polymer chains \cite{Su1980}.
The dispersion of such system contains two conduction bands, below and above the single-pillar energy, chosen as the zero reference.
The existence of the edge state in this case is determined by the Zak phase, which is an analog of the Berry phase\cite{Berry1984} defined on a unit cell of a size $d$ for a Wannier function $u_{nk}\left(x\right)$, integrated over a given band $n$ \cite{Zak1989}.
\begin{equation}
\zeta _n  = \int_{ - {\pi  \mathord{\left/
 {\vphantom {\pi  d}} \right.
 \kern-\nulldelimiterspace} d}}^{{\pi  \mathord{\left/
 {\vphantom {\pi  d}} \right.
 \kern-\nulldelimiterspace} d}} {\frac{{2\pi }}
{d}\int_0^d {u_{nk}^* \left( x \right)i\frac{{\partial u_{nk} \left( x \right)}}
{{\partial k}}dx} dk} 
\end{equation}

The Zak phase is determined by the ratio of the tunneling coefficients within and between the dimers  \cite{Delplace2011}: $\zeta_n=0$ if $t'/t>1$, and $\zeta_n=\pi$, if $t'/t<1$.
The topological transition $\zeta _n  = 0 \leftrightarrow \zeta _n  = \pi $ occurs at $t=t'$.
If the Zak phase is $\pi$, the number of states in the bulk is less than the number of pillars: $N=M-2$, and the remaining states, whose energy is that of uncoupled modes, are localized on the edges of the chain.
If the Zak phase $\zeta_n=0$, the number of states in the bulk is equal to the number of pillars $N=M$, and no edge states appear. The advantage of the optical systems is that the Zak phase can be measured directly \cite{Poddubny2015}.

In our system, as can be deduced from Fig. 1(a) and (b), a pair of edge states does exist in one polarization (anti-diagonal for our parameters), and does not exist in the other.
The result of the diagonalization of the Hamiltonian (1) with spin (see \cite{suppl}) for a chain of 30 pillars ($\delta J=0.3J$ for visibility) is shown in Fig. 1(c), with polarization shown with color. We see that the polarization states are interleaved, the lower band ends with polarization D, and the upper band begins with D, so the edge states (seen in the gap) are both necessarily A-polarized. Rashba SOC gives a similar result. 
For odd number of pillars, if $t/t'\neq 1$, one can redefine a dimer so that the Zak phase will be $\pi$ and a state will appear on the edge which contains a unit cell broken by the boundary.
For polaritons, an important consequence is that for one polarization (D), the edge state is on the right edge of the chain (Fig. 1(d)), and for the other polarization (A) the edge state is on the left edge of the chain (Fig. 1(e)). Calculation yields Fig. 1(f), where for all states, including the edge ones, the polarization is interleaved.

Overall, whatever the number of pillars in a finite zigzag chain, because of the polariton SOC there are always two edge states in the system, either having the same polarization when the number of pillars is even, or being cross-polarized when the number of pillars is odd.
The edge state polarization is always orthogonal to the axis linking the two last pillars at the chain edge.

\begin{figure}[h]\label{fig2}
\includegraphics[scale=0.35]{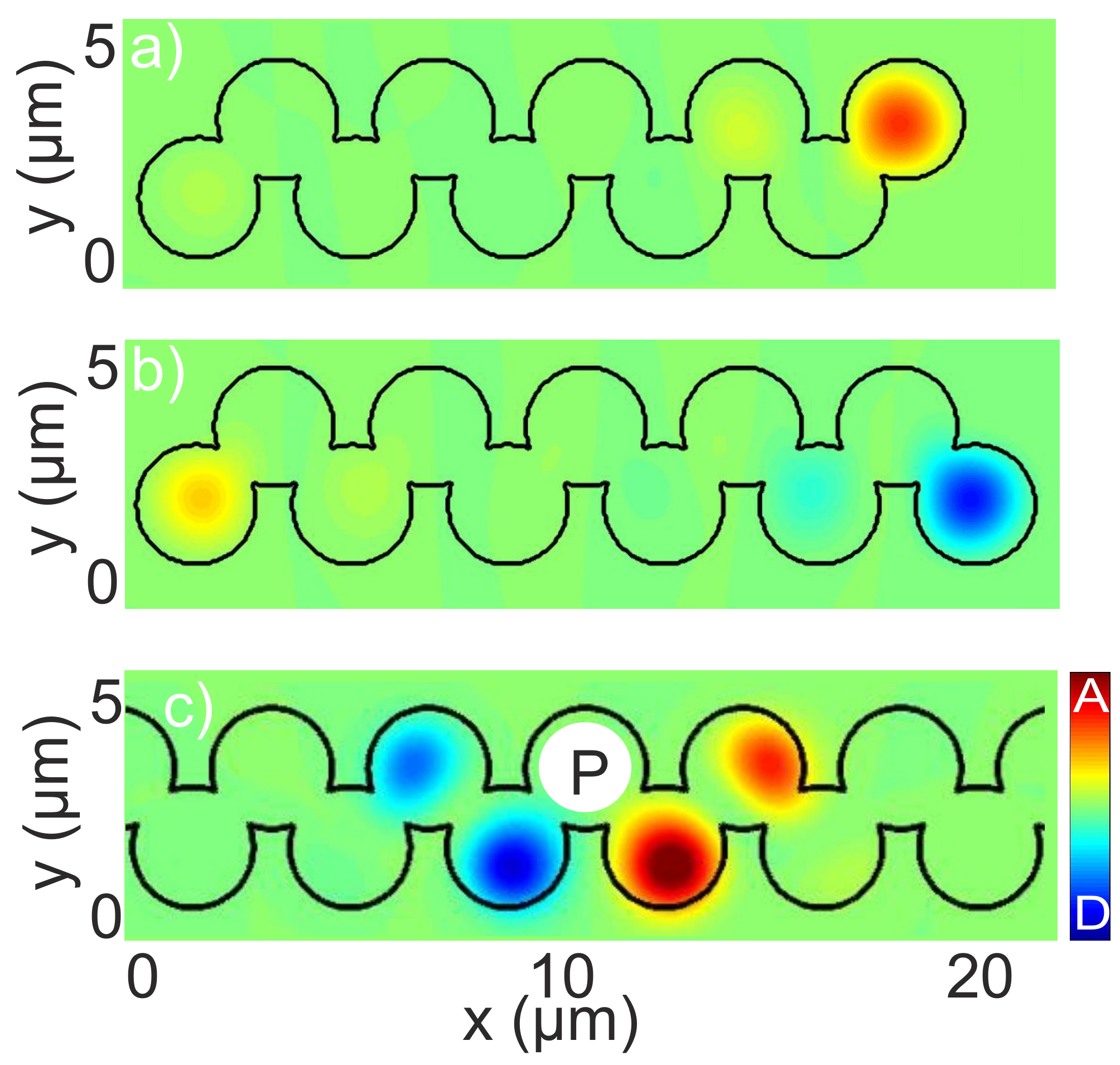} 
\caption{(color online) Calculated spatial images of the difference in diagonal polarization emission $I_D-I_A$: a) edge states in a chain with an even number of pillars; b) edge states in a chain with an odd number of pillars; c) Emission of the condensed states under localized pumping (marked P). Opposite diagonal polarization is observed on opposite sides.}
\end{figure}

We confirm the predictions of the analytical tight-binding model by solving numerically the spinor Schrodinger equation on a grid to find the eigenstates:
\begin{equation}
E\psi_\pm = -  \frac{{\hbar ^2 }}
{{2m}}\Delta \psi_ \pm + \beta {\left( {\frac{\partial }{{\partial x}} \mp i\frac{\partial }{{\partial y}}} \right)^2}{\psi_ \mp } + U\psi _ \pm,
\end{equation}
where $\psi(\mathbf{r},t)=({\psi_+(\mathbf{r},t), \psi_-(\mathbf{r},t)})^T$ are the two circular components of the wave function, $\beta ={\hbar ^{2}}\left( {m_{l}^{-1}-m_{t}^{-1}}\right) /4m$, while $m_{l,t}$ are the effective masses of TM and TE polarized particles respectively and $m=2\left( {{m_{t}}-{m_{l}}}\right) /{m_{t}}{m_{l}}$; $m_t=5\times10^{-5}m_0$, $m_l=0.95m_t$; $m_0$ is the free electron mass; $U(\mathbf{r})$ is the potential of the pillars describing the confinement of polaritons in the chain beyond the tight-binding model \cite{Nalitov2014b}.
The results of these calculations are presented in Fig. 2, showing the spatial images of the difference between the diagonal polarizations $I_D-I_A$ for both even (a) and odd (b) number of pillars. The localization length is discussed in supplemental material \cite{suppl}. 

\emph{Condensation on localized edge states}
A very effective way to excite these localized edge states is to create a polariton condensate using focused non-resonant pumping.
This technique allows creation of strongly out-of-equilibrium states, typically the states showing the best spatial overlap with the localised excitonic reservoir induced by the pump \cite{Galbiati2012, Tanese, Jacqmin2014}.
In lattices, the repulsive potential induced by the excitonic reservoir becomes attractive for particles with negative effective mass at the band edges, which leads to the condensation on localized gap states bound to the reservoir \cite{Tanese, Jacqmin2014}.
One expected peculiarity of the zigzag chain with a local pump is that the polarization of the localized mode where the condensation occurs is entirely fixed by the chain topology and the position of the pump, and does not rely on a symmetry breaking process.
To demonstrate this predicted feature, we model polariton condensation using the Hybrid Boltzmann-Gross Pitaevskii equation which includes relaxation mechanisms \cite{Jacqmin2014,Solnyshkov2014,Wertz2012}. For a thermal excitonic reservoir, the model can be reduced to:
\begin{eqnarray}
&i\hbar \frac{{\partial \psi _ \pm  }}
{{\partial t}}   =  - \left(1-i\Lambda\right) \frac{{\hbar ^2 }}
{{2m}}\Delta \psi _ \pm   + \beta {\left( {\frac{\partial }{{\partial x}} \mp i\frac{\partial }{{\partial y}}} \right)^2}{\psi _ \mp }   \\
&  + U\psi _ \pm -\frac{i\hbar}{2\tau}\psi_\pm
+\left(\left(U_R+i\gamma (n)\right)\psi_\pm+\chi\right) \exp \left(  - \frac{{\left( {{\mathbf{r}} - {\mathbf{r}}_0 } \right)^2 }}
{{\sigma ^2 }} \right) \notag
\end{eqnarray}
where the parameters (values from ref. \cite{Jacqmin2014}) other than in Eq.(3) are: $\Lambda$ -- the kinetic energy relaxation term, $U_R$ -- the reservoir potential amplitude, $\sigma$ -- the reservoir width, $\tau$ the lifetime, $\chi$ -- the Gaussian noise term included to describe the spontaneous scattering \cite{Stoof1999,Gardiner2002}, $\gamma(n)$ is the saturated stimulated scattering rate from the reservoir, $n$ is the total polariton density. We neglect the interactions within the condensate here.
A pump located close to the edge will excite the unique localized mode at the edge of the chain.
If the pumping spot is located in the bulk, the potential of the reservoir cuts the lattice into two smaller chains, and the same reasoning as above applies to each of them, leading to the condensation at their respective edge states. The results of the simulations for the pumping spot located in the middle of a chain (which allows to check all predictions simultaneously) are presented in Fig. 2(c), showing the difference between the intensities of the diagonal polarizations $I_D-I_A$ of the light emitted from the system above the condensation threshold. The condensation indeed occurs on the localized edge states on both sides of the spot, with polarization controlled by the condition on the Zak phase $\zeta_n=\pi$.

\begin{figure}[h]\label{fig3}
\includegraphics[scale=0.44]{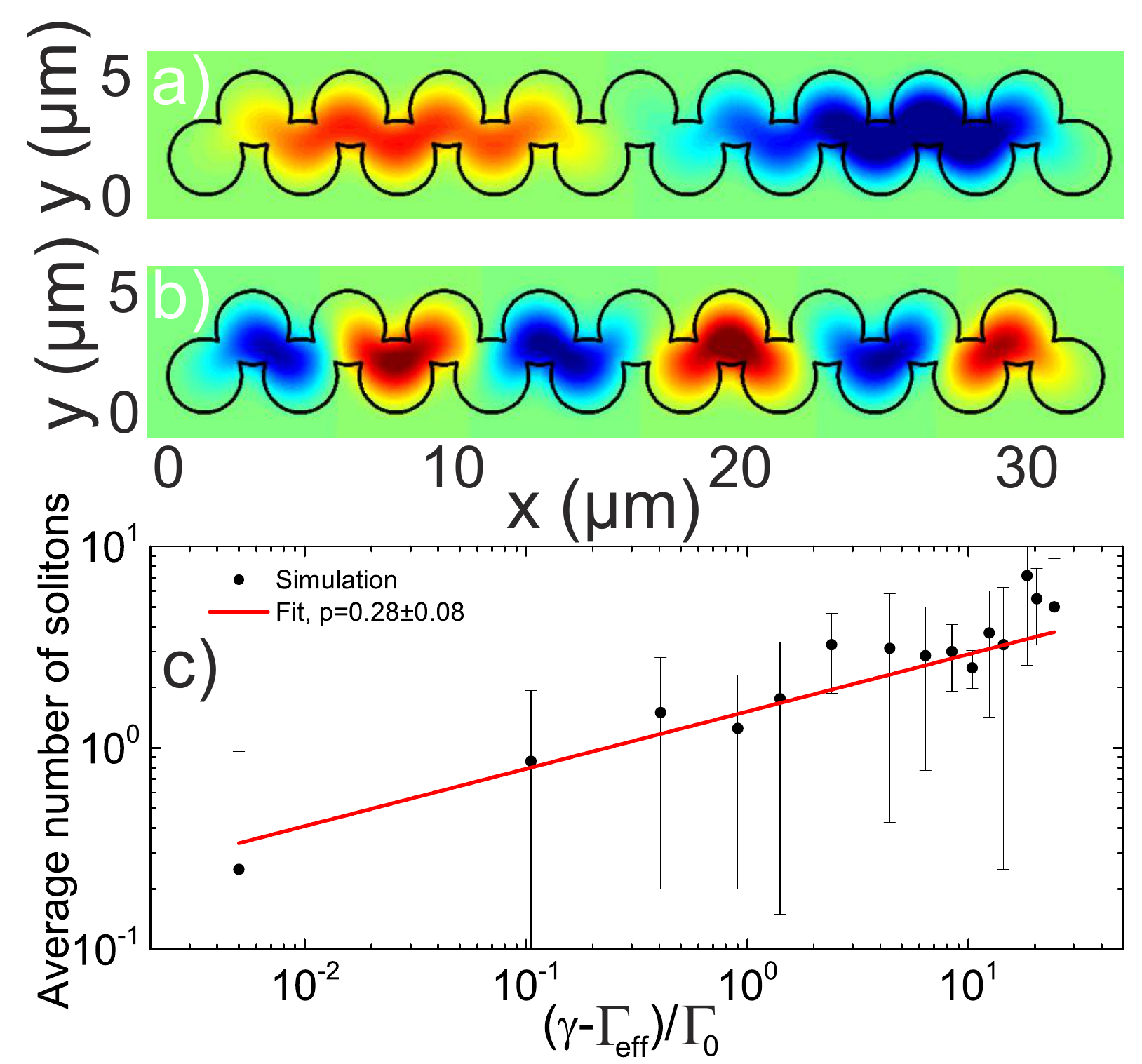} 
\caption{(color online) a,b) Difference in diagonal polarization intensities for 2 examples of polarization textures in a zigzag chain. Colors as in Fig. 2. c) Average number of solitons as a function of effective pumping fitted by power law.}
\end{figure}

\emph{Spontaneous formation of dark-bright solitons via the Kibble-Zurek mechanism}
The consequences of the non-trivial topology of the system are truly revealed under \emph{homogeneous} non-resonant pumping. While the previous results could be verified by purely linear measurements \cite{Poddubny2015}, in this part we study the condensation of  polaritons, accompanied with the emergence of a spinor order parameter.
As any second-order phase transition occurring on a finite timescale, it is described in terms of parameter quenching, responsible for the KZM-type formation of topological defects \cite{Kibble1976,Zurek1985,Lamporesi2013}.
In a spinor system, the large phase fluctuations at the early stage of a condensation process are associated with spatial fluctuations of the polarization, governed by the phase difference between the spin components.
The polarization domains correspond to the dimerization domains in the SSH picture, and the domain wall between them is equivalent to the SSH soliton \cite{Su1980}.
These dark-bright solitons separate two polarization domains, both equally stable and characterized by a fixed difference $\pi$ in their Zak phase. 
The stability of a dark-bright soliton, which cannot be destroyed by acceleration, is its most important feature: once formed, it does not disappear, contrary to a scalar dark or grey soliton. This stability requires the interactions in the condensate to be smaller than the TE-TM splitting \cite{suppl}. It allows to detect such objects in $cw$ experiments, being advantageous with respect to the previous proposals of the KZM studies with polaritons \cite{Matuszewski2014,Liew2015}, requiring single-pulse experiments. 

Fig.3 shows the results of simulations based on the numerical solution of the equation (4) with a \emph{homogeneous} reservoir potential ($\sigma=\infty$) and interactions \cite{suppl}.
Panels (a,b) show the difference between the intensities of the diagonal polarizations $I_D-I_A$ of the light emitted by a condensate formed under weak and strong pumping respectively. In panel (a), two domains A- and D-polarized corresponding to a single dark-bright soliton are visible.
Panel (b) shows 6 polarization domains and 5 domain walls (for a movie see supplementary \cite{suppl}).
In our numerical experiment, we do not change the temperature of the system, as in the classical KZM, but rather turn on the pumping and fill the system with particles, changing the critical condensation temperature, but keeping the system temperature (controlled by $\chi$ and $\Lambda$) constant. The quenching time is controlled by the pumping intensity $\tau_Q^{-1}\propto P$.
This scheme, while being simpler and ubiquitously present in all polariton experiments, fits the KZM scheme, because the relative temperature $\epsilon=(T-T_c)/T_c$ at threshold changes linearly with time (see \cite{suppl}).

Indeed, the  Gaussian noise $\chi$ is uncorrelated in time, and so creates a frequency-independent population of particles. However, the energy relaxation term proportional to $\Lambda$ describes energy-dependent decay acting on these particles. The resulting spectral density $|\psi (E)|^2$ for the polariton state of energy $E$ can be obtained as: $\left|\psi (E)\right|^2 \propto \chi/\Gamma$, where $\Gamma$ is the total decay rate, composed of  energy-independent $\Gamma_0$ (ground state lifetime) and energy-dependent relaxation $\Gamma_\Lambda=\Lambda E$ \cite{Solnyshkov2014}, giving:
\begin{equation}
\left|\psi (E)\right|^2 \propto \frac{\chi}{{{\Gamma _0} + \Lambda E}} \approx \frac{\chi}{{{\Gamma _0}}}\left( {1 - \frac{{\Lambda E}}{{{\Gamma _0}}}} \right)
\end{equation}
The linear part of the spectrum at low energies can be interpreted as a Boltzmann distribution function with an effective temperature $T=\Gamma_0/\Lambda$. Varying the relaxation efficiency, we can change this effective temperature: the better is the relaxation, the lower is the temperature.
According to KZM, the average density of the topological defects $n_{sol}$ in a BEC scales as the inverse healing length $\xi^{-1}=\xi_{0}^{-1}\left|\epsilon\right|^\nu$, where $\nu$ is a scaling exponent. In the mean-field approximation, $\nu=1/2$ (because $\xi=\hbar/\sqrt{2\alpha n m}$, where $\alpha$ is the interaction constant), and the dynamical exponent is $z=2$ \cite{Zurek1996,Damski2010} for relaxation linear in energy, which allows to write $n_{sol}=\xi_{0}^{-1}\left(\tau_0/\tau_Q\right)^{\nu/(1+z\nu)}\propto\left(\tau_0/\tau_Q\right)^{1/4}$. Thus, $n_{sol}\propto\tau_Q^{-1/4}\propto \gamma^{1/4}$: a scaling exponent of $1/4$ is expected for the mean-field universality class. 
Fig. 3(c) shows the number of solitons appearing in a chain of 40 pillars versus the effective pumping intensity $(\gamma-\Gamma_{eff})/\Gamma_0$, where $\Gamma_{eff}\approx 5.5 \Gamma_0$ is the effective decay rate accounting for $\Lambda$. Each point is obtained as an average of 10 simulations. The power law fit is compatible with the scaling exponent $1/4$ expected for  KZM.

To conclude, we have demonstrated that the Zak phase plays a crucial role for the description of condensation in 1D zigzag chains of polariton pillars. Because of the SOC, such chains always exhibit exponentially localized edge states. Similar results can be obtained for Rashba or Dresselhaus SOC in atomic BECs. Under homogeneous pumping, dark-bright solitons appear between the domains of orthogonal polarization via the Kibble-Zurek mechanism. We extract numerically the dependence of the soliton density against the quenching parameters and find it in agreement with the analytical predictions.
These domain walls can also be created by quasi-resonant excitation, for example, using Gauss-Laguerre beams focused on a pillar chain (above the bistability threshold) \cite{FlayacNJP}. They can also be manipulated using the electrically controlled in-plane effective magnetic fields \cite{Solnyshkov2013}, which might allow to design optical race-track memories \cite{Fert2013}.

We acknowledge discussions with M. Glazov, A. Amo, and J. Bloch and the support of ITN INDEX (289968) and ANR Labex GANEX (207681).

\section{Supplemental Material}

\subsection{Tight-binding Hamiltonian with polarization}

The matrix of the tunneling couplings in a single zigzag chain in the tight-binding approximation taking into account the spin-dependent tunneling is given by

\begin{equation} \label{Hamiltonian}
\mathrm{F}_{\mathbf{k}} = - \left( \begin{matrix}
f_{\mathbf{k}} J & f_{\mathbf{k}}^+ \delta J \\
f_{\mathbf{k}}^- \delta J & f_{\mathbf{k}} J
\end{matrix} \right),
\end{equation}
where complex coefficients $f_{\mathbf{k}}$,$f_{\mathbf{k}}^\pm$ are defined by:
\begin{equation}
f_{\mathbf{k}}= \exp(-\mathrm{i}\mathbf{k d}_{\varphi}),\quad
f_{\mathbf{k}}^\pm =  \exp(-\mathrm{i}\left[\mathbf{k d}_{\varphi} \mp 2 \varphi \right]), \notag
\end{equation}
and $\varphi = \pm \pi / 4$ is the angle between horizontal axis and the direction to the $j$th nearest neigbor. Each pillar in a chain has only 2 neighbours, contrary to the planar graphene case. $J$ is the polarisation independent tunneling coefficient, whereas $\delta J$ is the SOI-induced polarization dependent term. The results of the diagonalization of the Hamiltonian are shown in Fig. 1(c,f) of the main text, where the edge states  appear in between the two dispersion bands. The polarization degree of the bulk states on this figure is smaller in the case of odd number of pillars, because  the extra pillar affects the overall equivalence of the two diagonal polarizations.

\subsection{Localization length of the edge states}

The inverse localization length $\kappa$ can be estimated by requiring that the corresponding wavefunction vanishes on the opposite edge:
\[
t'\sinh \kappa \left( {M + 1} \right)a = t\sinh \kappa Ma
\]
where $a$ is the length of the unit cell, and M is the number of pillars in the chain.
In the limit of sufficiently large difference of $t$ and $t'$, the equation reduces to $ t'/t \approx \exp \left( { - \kappa a} \right)$
which means physically that we deal with a particle whose mass is given by the larger of the coefficients $t$, which decays inside a barrier of a height $t'$ (which is the energy gap between the two bands).

The numerical simulations also confirm that the decay of the edge state is indeed exponential, as can be seen in Fig. S3, where the numerical simulations (using the 2D model described in the main text) are compared with the analytical expression obtained in the framework of the tight-binding approximation. The good quality of the approximation in this case is due to the direct tunneling of photons within the confinement potential.

\begin{figure}[h]\label{figs1}
\includegraphics[scale=0.3]{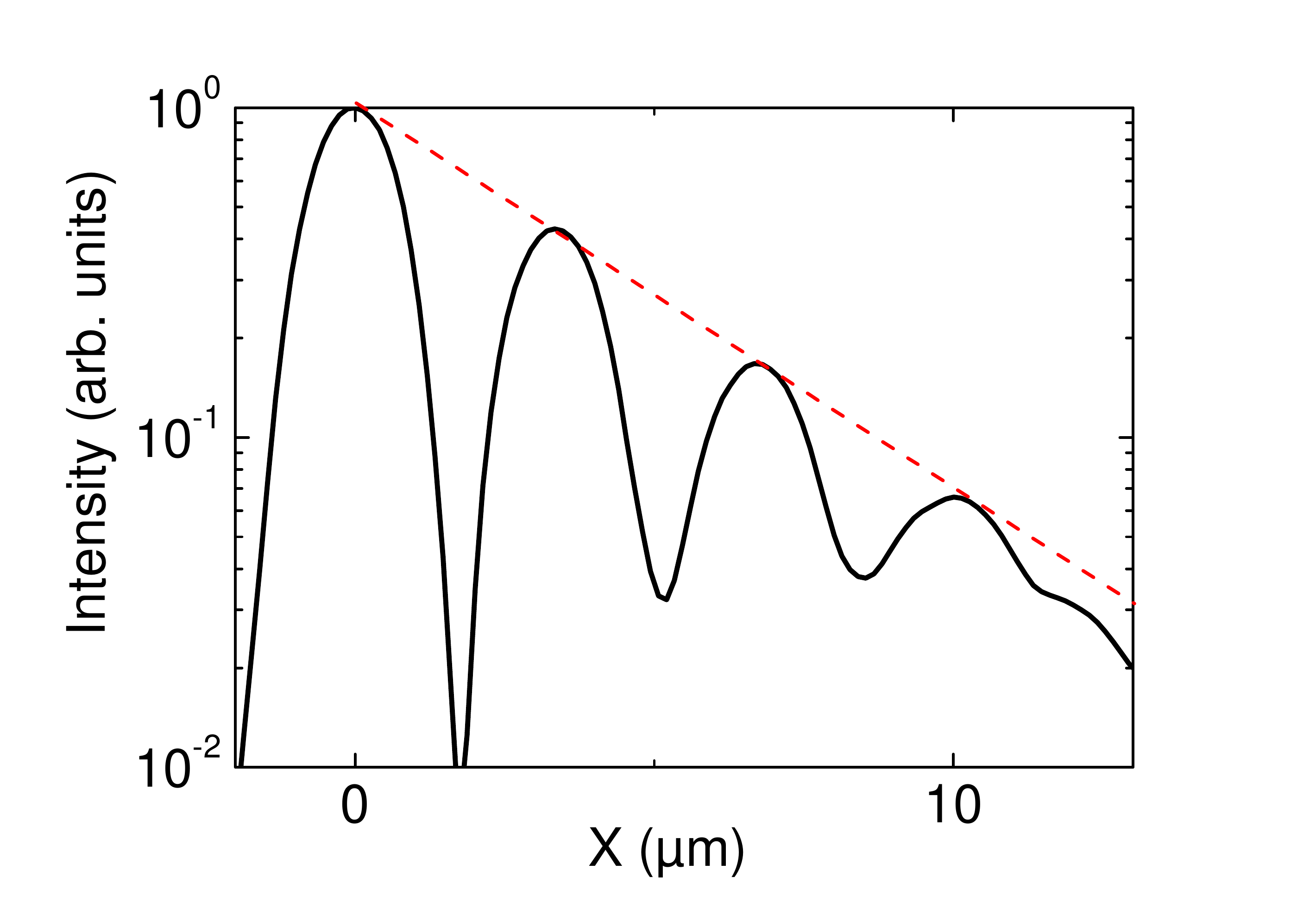} 
\caption{(color online) The exponential decay of the edge state: solid line -- numerical simulation, dashed line -- analytical solution.}
\end{figure}

The exponential localization of the edge states plays a very strong role for the condensation on these states in case of localized pumping, as explained in the main text. In case of pumping on the edge, the localized edge states, which always exist in polariton zigzag chain, as shown above, will  have a much larger overlap with the reservoir than any of the propagative bulk states, with the overlap ratio being $L/l$, where $L$ is the length of the chain and $l=\kappa^{-1}$ is the localization length of the edge state, discussed above.

\subsection{Formation of a condensate and the Kibble-Zurek mechanism}

In this section we discuss the simulations and the corresponding experiments required for the study of the topological defect formation via the Kibble-Zurek mechanism under non-resonant pumping. The system is composed of a homogeneously pumped exciton reservoir and polariton states.  As written in the main text, we describe the system by the spinor Gross-Pitaevskii equation with phenomenological terms describing decay, energy relaxation, stimulated scattering from the reservoir into the condensate, and the noise, describing the spontaneous scattering:

\begin{eqnarray}
&i\hbar \frac{{\partial \psi _ \pm  }}
{{\partial t}}   =  - \left(1-i\Lambda\right) \frac{{\hbar ^2 }}
{{2m}}\Delta \psi _ \pm   + \beta {\left( {\frac{\partial }{{\partial x}} \mp i\frac{\partial }{{\partial y}}} \right)^2}{\psi _ \mp }   \\
& +\alpha \left|\psi_\pm\right|^2\psi_\pm + U\psi _ \pm -\frac{i\hbar}{2\tau}\psi_\pm
+\left(U_R+i\gamma (n)\right)\psi_\pm+\chi \notag
\end{eqnarray}

Here, $\alpha$ is the interaction constant for polaritons having the same spin. For opposite spins, the interactions pass through the dark exciton states, and the corresponding constant is negligibly small, as a second-order correction. Thus, the interactions are strongly spin-anisotropic, which allows to consider the two spin components as being almost decoupled during their formation. The stimulated scattering term $+i\gamma(n)\psi$ in the Gross-Pitaevskii equation leads to the exponential increase of the particle density with time, until the saturation is reached. The saturation is obtained by the inclusion of exponential saturation factor $\gamma(n)=\gamma_0 \exp(-|\psi|^2/n_{max})$. It is the constant prefactor $\gamma_0$ which determines the speed with which the condensation threshold is passed $\epsilon(t)=(T-T_c)/T_c(t)$, and therefore, the number of topological defects via the Kibble-Zurek mechanism. In this type of experiment, it is not the system's effective temperature $T$, but the critical temperature $T_c$ which changes with time, while the system passes the condensation threshold. In 1D, $T_c\propto n^2$, where $n\propto \exp(\gamma t)$, which, after the linearization of the exponent, gives at threshold a linear behavior $\epsilon(t)=t/\tau_Q$, where $\tau_Q\propto \Lambda/2\Gamma_0\gamma$, allowing to test the scaling behavior as a function of $\gamma$.

In experiments, it is the pumping intensity which controls the reservoir density, which, in turn, controls the stimulated scattering efficiency. The dominant relaxation mechanism is usually the exciton-exciton scattering, and thus the probability for the reservoir particle to scatter into the condensate is proportional to $W_0 (1+N_{cond})N_{res}^2$, where $W_0$ is a constant prefactor proportional to the exciton-exciton interaction strength. Varying the reservoir density by changing the pumping, one can therefore change the stimulated scattering rate $\gamma_0\propto N_{res}^2$ in experiments. Of course, this also changes the saturation value for the condensate density, but the Kibble-Zurek mechanism depends only on the parameters important at the transition point. We have specifically checked in numerical simulations, where the parameters can be changed independently, that the saturation density $n_{max}$ does not affect this mechanism, if the density remains sufficiently small for the TE-TM coupling to remain dominant over the interactions. If the interactions dominate the spin-orbit coupling, the topological gap closes, the dark-bright solitons are replaced by less stable half-solitons, and, finally, a homogeneous linearly polarized condensate, corresponding to the ground state of the system, emerges. In our simulations such behavior occurs if the pumping intensity is increased significantly above the range shown in Fig. 3 and Fig. S2.

 The behavior of $\Lambda$ as a function of the reservoir density also has to be taken into account to describe the experiments properly. If a linear dependence is assumed for exciton-exciton relaxation mechanism, $\tau_Q^{-1}\propto \gamma/\Lambda\propto P$, where $P$ is the pumping of the reservoir, which means that dependence on $P$ in experiments can be expected to be the same as the dependence on $\gamma$ in numerical simulations. 

\subsection{Additional results on KZM}
In this section we present additional results with KZM simulations. We have compared the formation of topological defects for two different values of the relaxation constant $\Lambda$, determining the quench time $\tau_Q\propto\Lambda$. Fig. S2 shows that, as expected, a faster quench (smaller $\Lambda$) leads to a higher number of solitons, but the power law remains the same, with a scaling exponent $1/4$. A smaller system size of 19 pillars, as compared to Fig. 3c) of the main text (40 pillars), was used for these simulations, in order to shorten the calculation time.

\begin{figure}[h]\label{figs1}
\includegraphics[scale=0.3]{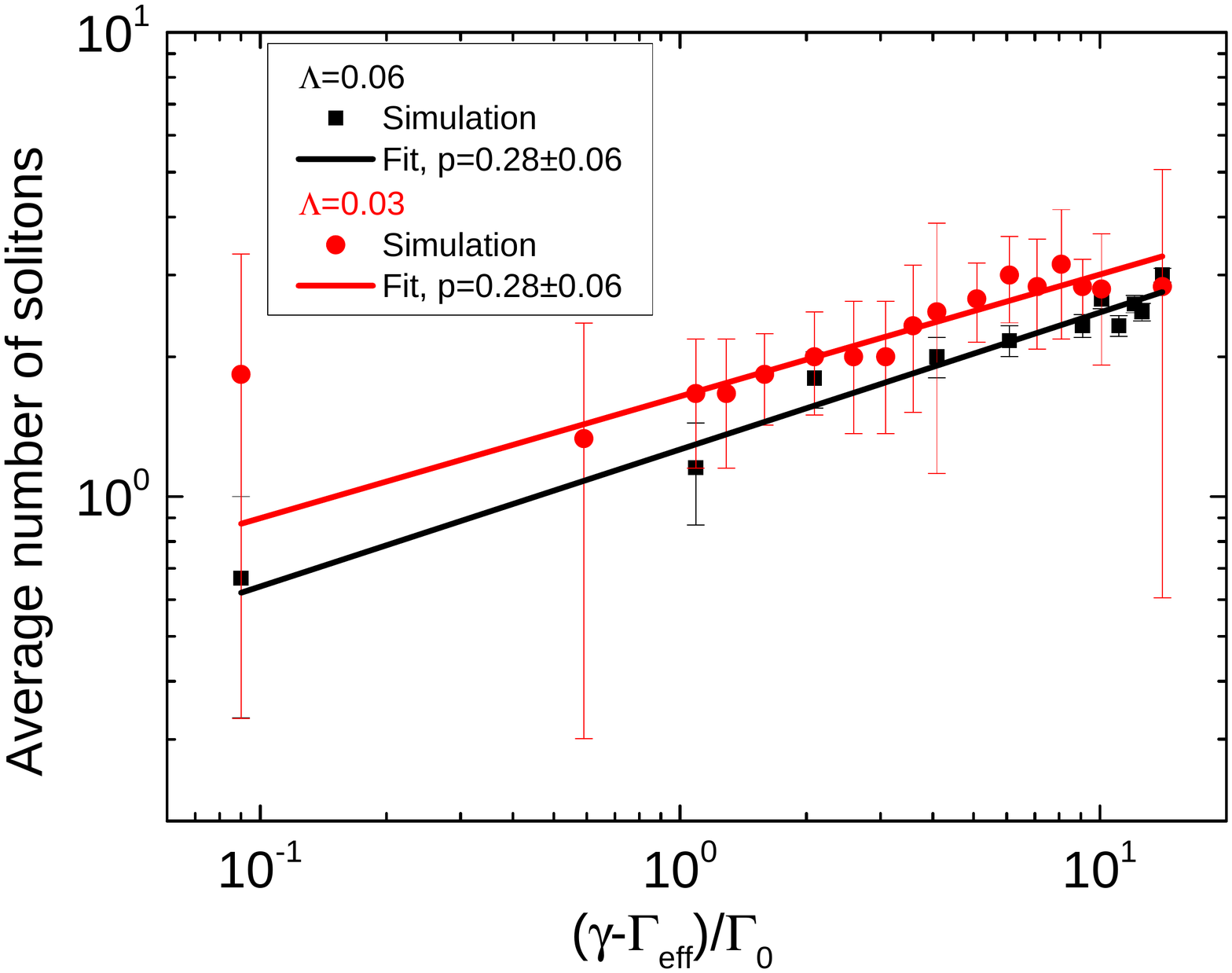} 
\caption{(color online) Average number of solitons as a function of effective pumping for different relaxation efficiency: $\Lambda=0.06$ (black) and $\Lambda=0.03$ (red). Power law fits are shown with solid lines. }
\end{figure}

We note that both in the main text and in the supplemental material, the fit gives a slightly higher scaling exponent of about 0.28 instead of 0.25. Current precision of the simulations does not allow to distinguish between the two values, but future studies might show if this deviation from the analytical predictions is significant, and if it is due to the particular (for example, topological) properties of the system.

Finite-size effects take place in our system for higher values of pumping, when the distance between the solitons becomes comparable with the period of the chain. In this case, topological protection is reduced, and solitons recombine, which leads to their average number being much smaller than the one predicted by scaling law. These effects begin approximately at $\gamma/\Gamma_0=20$ for 19-pillar chain and at $\gamma/\Gamma_0=35$ for a 40-pillar chain. This is why it is difficult to obtain a large number of solitons in the chain.

\section{Video}
The supplemental Video file demonstrates the polariton condensation in a zigzag chain of micropillars. The colormap corresponds to the difference between the intensities of the diagonal polarizations, $(I_D - I_A)$, as in the figures 2 and 3 of the main text. One can observe the formation of several polarization domains and the establishment of a steady state.

\bibliography{reference}

\end{document}